\documentclass[3p, 10pt]{elsarticle} % options: [preprint], [review], [1p], [3p,twocolumn], [5p,twocolumn], [authoryear]
 
\usepackage{lipsum}
\makeatletter
\def\ps@pprintTitle{%
 \let\@oddhead\@empty
 \let\@evenhead\@empty
 \def\@oddfoot{}%
 \let\@evenfoot\@oddfoot}
\makeatother
\usepackage{algorithm}
\usepackage{algpseudocode}
\usepackage{amsmath}
\makeatletter
\let\oldtheequation\theequation
\renewcommand\tagform@[1]{\maketag@@@{\ignorespaces#1\unskip\@@italiccorr}}
\renewcommand\theequation{(\oldtheequation)}
\makeatother

\usepackage{amssymb} % math symbols
\usepackage{caption}
\usepackage{color}
\usepackage{epstopdf}
\usepackage{float} 
\usepackage{graphicx} % include figures in document
\usepackage{lmodern}
\usepackage{lineno} % numbers in line
\usepackage{lscape} % use landscape in documents
\usepackage{mathtools}
\usepackage[numbers]{natbib}% natbib for els article
\usepackage{nicefrac}
\usepackage{setspace}
\usepackage{subcaption}
\usepackage{wrapfig} 
\usepackage{xfrac}
\usepackage[font=footnotesize,labelfont={bf,sc,small},figurename=Fig.,tablename=Tab.]{caption}
\usepackage[colorlinks=true,linkcolor=blue,citecolor=blue]{hyperref} 

\newcommand{\e}{\mathrm{e}}

\begin{document}
\bibliographystyle{elsarticle-harv.bst}\biboptions{authoryear}

\begin{frontmatter}
\title{Geological modeling using Recursive Convolutional Neural Networks}
\author[Queens]{Sebasti\'an  \'Avalos\corref{cor1}}
\ead{sebastian.avalos@queensu.ca}
\author[Queens]{Juli\'an M. Ortiz\corref{cor1}}
\ead{julian.ortiz@queensu.ca}
\address[Queens]{The Robert M. Buchan Department of Mining, Queen's University, Canada}
\cortext[cor1]{Corresponding authors}
%\date{}
% Abstract
\begin{small}
\begin{abstract}
\indent
Resource models are constrained by the extent of geological units that often depends on lithology, alteration and mineralization. Three dimensional geological units must be built from scarce information and limited understanding about the geological setting. In this work, a new technique for Multiple-Point Geostatistics simulation based on a Recursive Convolutional Neural Network approach (\textsc{RCNN}) is presented. The method requires conditioning data and a training image. The training image is used to learn geological patterns and simulation is carried out by using the previous conditioning data.

A lithological modeling process is carried out in a synthetical 3D surface-based model lithology to demonstrate the method. Spatial metrics are used to measure performance and characterize the RCNN properties. Also, strengths and weaknesses of the methodology are discussed by briefly reviewing the theoretical perspective and looking into some of its practical aspects.

\end{abstract}

\end{small}

\end{frontmatter}

\section{Introduction}
Multiple-point statistics (MPS) simulation methods provide a framework for spatial modeling of variables in geoscience problems. The reproduction of complex patterns and spatial uncertainty assessment are their main goals. They can be applied for continuous or categorical variables and the methods can be extended to the multivariate case, suiting applications in diverse fields (oil and gas, hydrological reservoirs, ore deposits). 

MPS simulation has seen an explosive development since Strebelle's seminal paper in 2002 \citep{strebelle2002conditional}. Many different approaches have been proposed \citep{arpat2007patterns, mariethoz2010direct, parra2011texture} and some have reached industrial applications \citep{mariethoz2010direct}. The methods associated to MPS are intimately related to image and texture analysis \citep{daly2004highorder, parra2011texture, zahner2015image}. Excellent reviews of multiple-point simulation methods are available \citep{tahmasebi2018multiple} and further theoretical and practical details can be found \citep{mariethoz2014multiple}. 

This paper is motivated by the possibilities offered by deep learning methods, particularly Convolutional Neural Networks (CNNs) \citep{lecun1998gradient} which have shown extraordinary performances in image analysis, including classification, reconstruction, segmentation, labeling and more \citep{dosovitskiy2015learning, eilertsen2017hdr, gatys2016image, he2016deep}. The main aim is to create a bridge between CNNs and MPS. This is done by proposing a Recursive Convolutional Neural Network (\textsc{RCNN}) approach that enables the architecture to learn features of an underlying phenomenon from a training image and then perform simulations on different domains reproducing the phenomenon complexity, conditioned by hard data at sample locations.

\section{Convolutional Neural Networks} \label{RelWork}

CNNs belong to a set of deep learning architectures with great abilities on image analysis. Their features extraction properties, from raw images, have made them suitable for a variety of image analysis applications. Connecting neural networks with MPS is not new \citep{caers1998stochastic} however deep learning techniques have only recently been connected to MPS \citep{laloy2018training}. 

CNNs have the main feature of sharing inner parameters across the network, which leads to architectural properties of scale, shift and distortion invariance, making them a powerful tool for image feature extraction. Those properties mean that regardless where and how a specific raw feature appears in the image, a suitable and well-trained CNN is able to capture that feature. Once features are extracted, images can be classified, segmented or even reconstructed. 

\begin{figure}[H]
\begin{center}
\includegraphics[width=1\textwidth]{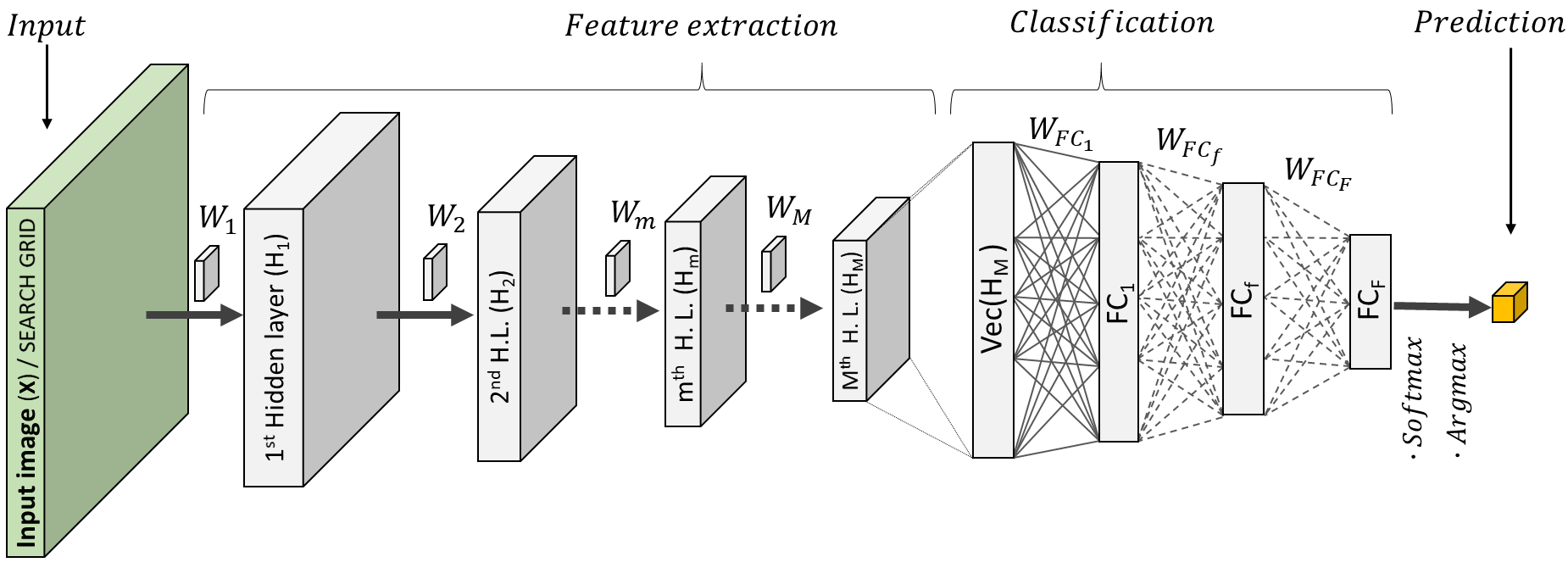}
\caption{CNN architecture showing the features extraction and classification zones together with main notations.}
\label{figSimpleConvNet}
\end{center}
\end{figure}

CNNs are composed by a \emph{feature extraction} block and a \emph{classification} block (\autoref{figSimpleConvNet}). The first block receives a grid-like topology input and extracts representative features in a hierarchical manner. The second block receives the top hierarchical feature and delivers a final matrix of prediction. 

From now on, a three dimensional image grid-like topology is used to depict the inner process of a CNN. The building blocks of the feature extraction section are: input image $\mathbf{X}$, filters $\mathbf{W}_m$, bias vectors $\mathbf{b}_m$ and hidden layers $\mathbf{H}_m$. Then, $\mathbf{H}_m$ results of convolving $\mathbf{W}_m$ over $\mathbf{H}_{m-1}$, adding a bias vector $\mathbf{b}_m$ and then passing the temporary result through (1) a Batch-Normalization function, (2) a non-linear activation function and (3) a pooling function as:

\begin{equation}
\small
\label{HiddenLayers}
\mathbf{H}_m =
\begin{cases}
\mathbf{X} & \text{if }m = 0\\
pool(g(BN(\mathbf{W}_m\mathbf{H}_{m-1}+\mathbf{b}_m))) & \text{if } 0 < m \leq M
\end{cases}
\end{equation}

A convolution is understood as the process of sliding the filter over the input while performing the sum of an element-wise multiplication between the filter values and the corresponding section of the input. 
Activation functions, $g(\cdot)$, a set of non-linear transformations, are required to suitably extract features. They are applied element-wise over each feature map in the hidden layers. Some activation functions are \emph{Sigmoid}, \emph{TanH} and \emph{ReLU}. Each activation function has an active zone, that is, an interval where the derivative of the function is not zero. 

The Batch Normalization function (BN) \citep{ioffe2015batch} normalizes $ \big  ( \mathbf{W}_m\mathbf{H}_{m-1}+\mathbf{b}_m \big )$ before passing it through the activation function by subtracting the mean and dividing by their respective standard deviation over each feature map. This process has three main properties: (1) it avoids vanishing gradient problems during training, by adjusting the input values to the active zone; (2) it accelerates the training process; and (3) it serves as a regularization method. 

The pooling function, $pool (\cdot)$, seeks to reduce the size of the representative hidden layer by taking small regions of each feature map. Several functions exist including the widely used  \emph{max pooling}, and others such as \emph{average pooling}, \emph{min pooling} and \emph{L2-norm pooling}. They are usually applied to go from a moving window of $(2,2,2)$ to a single value $(1,1,1)$. \emph{Max pooling} keeps only the maximum value among the nodes in the small region. This pooling function significantly reduces the number of learning parameters, improving statistical efficiency and reducing the memory storage consumption \citep{goodfellow2016deep}.

The classification section is composed by a set of feedforward networks \citep{hornik1989multilayer} whose building blocks are: weight matrices $\mathbf{W}_{FC_f}$, bias vectors $\mathbf{b}_f$ and hidden fully connected layers $\mathbf{FC}_f$. Then, $\mathbf{FC}_f$ results of a matrix multiplication between $\mathbf{W}_{FC_f}$ and  $\mathbf{FC}_{f-1}$, adding a bias vector $\mathbf{b}_f$ and then passing the temporary result only by (1) a Batch-Normalization function and (2) an activation function as:

\begin{equation}
\small
\label{FullyConnectedFinal}
\mathbf{FC}_f =
\begin{cases}
Vec(\mathbf{H}_M) & \text{if } f = 0\\
g(BN(\mathbf{W}_{FC_f} \mathbf{FC}_{f-1}+\mathbf{b}_f) )& \text{if } 0 < f \leq F
\end{cases}
\end{equation}

The first feedforward network $\mathbf{FC}_0$ corresponds to a vector-representation $Vec(\mathbf{H}_M)$ of the last hidden layer in the feature extraction block. The BN and activation functions act exactly as presented before. 

The final layer $\mathbf{FC}_F$ must have the same dimensions as the number of categories when categorical distributions are required.. Let $K$ be the number of categories, $n_{FC}^{(F)}=K$, and $s_k \in \{s_1, ..., s_K \}$ the non-normalized conditional probability of each class in $\mathbf{FC}_F$. By passing $\mathbf{FC}_F$ through a softmax function:

\begin{equation}
\small
\label{SotfmaxSimple}
 \hat{p}(k) = \dfrac{\e^{s_k}}{\sum_{c=1}^K \e^{s_c} }
\end{equation}
\noindent
the expected conditional probabilities for each class are obtained. Then, the final predicted state $\hat{s}_k$ is inferred by taking the position $k$ which contains the highest probability, that is, using the argmax function.
% %%%%%%%%%%%%%%%%%%%%%

Inner parameters $\Theta = \{ \mathbf{W},\mathbf{b} \}$ are initialized as random values from a normal gaussian truncated function. Optimum $\Theta$ values are obtained as result of training the CNN by applying the Adam optimizer \citep{kingma2014adam} algorithm. The loss functions to minimize during training (\autoref{eqLossFunctions}) for categorical variables, considering the predicted probability of each class (\autoref{SotfmaxSimple}) and the real probability $p\big(k)$, is the negative log-likelihood of the Cross Entropy ($CE$):
\begin{equation}
\small
\label{eqLossFunctions}
	CE: \mathcal{L}(\Theta) = - \sum_{k=1}^K \,\, \big( \log{ \hat{p}(k)} \big) \cdot  p ( k ) 
\end{equation}

% %%%%%%%%%%%%%%%%%%%
% %%%%%%%%%%%%%%%%%%%
% %%%%%%%%%%%%%%%%%%%
% %%%%%%%%%%%%%%%%%%%

\section{The Recursive Convolutional Neural Network approach} \label{methoSection}

\begin{figure}[H]
\begin{center} 
\includegraphics[width=1\textwidth]{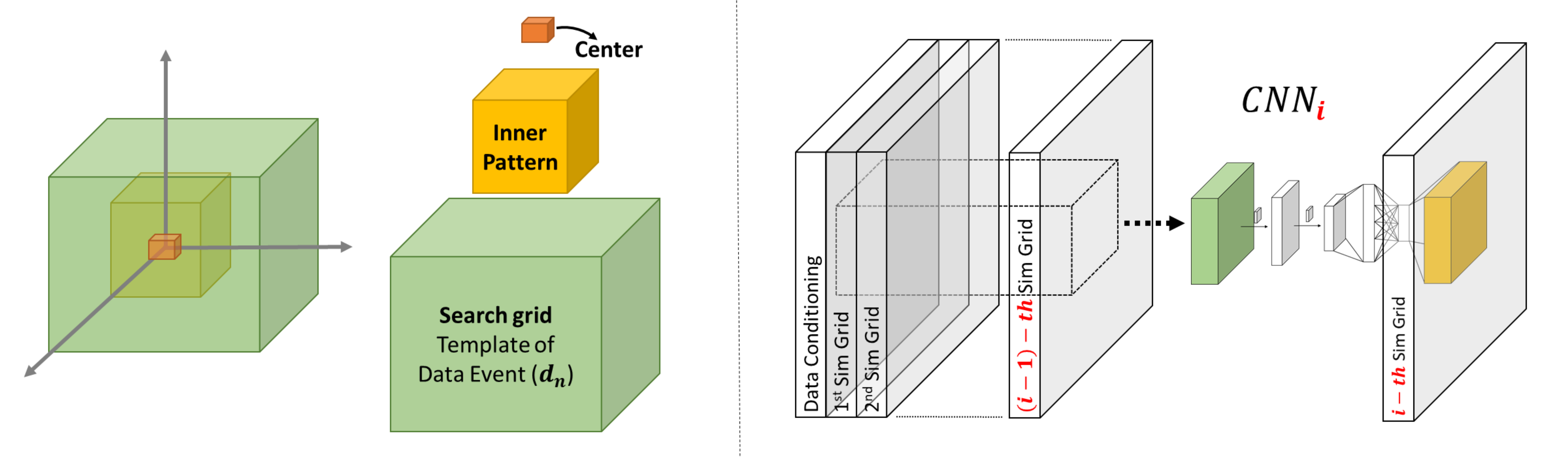}
\caption{(Left) Illustration of Search Gird and Inner Pattern concepts. (Right) Scheme of the CNN$_i$ in a RCNN architecture.}
\label{figRCNNarchitecture}
\end{center}
\end{figure}

Let SG and IP be the search grid and inner pattern, whose dimensions are odd positive integers to ensure the existence of a collocated center (\autoref{figRCNNarchitecture}. Left). In MPS terms, the SG is the neighbourhood (template) that contains the data event $d_n$ (conditioning data).

The main idea is that a first CNN$_1$ is trained to predict a conditional cumulative distribution function (ccdf) of each location (node) inside an IP by receiving, as input, a data event ($d_n$) embedded in a SG (\autoref{figRCNNarchitecture}. Left). Then a second CNN$_2$ is trained to predict the same ccdf but now knowing the same input and the IP predicted by the previous CNN$_1$. The third CNN$_3$ is trained using same input and the two previously predicted IP by CNN$_2$ and CNN$_1$. The process can be done again (\autoref{figRCNNarchitecture}. Right) giving the \emph{recursiveness} property to the RCNN approach. The idea behind the recursiveness is the improvement on results quality by taking into account the previously simulated information. 

The \textsc{RCNN} training algorithm starts by creating $N+1$ simulation grids (domains), namely $D^0, D^1,...,D^N$, whose dimensions are equal to the TI and then extracting a percentage of hard data ($perDC$) randomly from the TI and assigning them to all $D^i$. The simulation process, explained later, is carried. Once all $D^i$ are simulated, every pair-list database (DB$^i$) of input-output as $\mathbf{X}^i \leftrightarrow$ IP$^{Real}$ is created in order to train the respective CNN$_i$. This is done by extracting $\forall i$ $SG(D^i)$ in order to create $\mathbf{X}^i$ and the respective collocated IP$^{Real}$. 

Using the CE of \autoref{eqLossFunctions}, the loss function of each CNN$_i$ is:

\begin{equation}
\small
\label{ParticularLossFunction}
	\mathcal{L}_i(\Theta^{i}) = -\sum_{a=1}^{ip_x} \sum_{b=1}^{ip_y} \sum_{c=1}^{ip_z} \sum_{k=1}^K \,\, \Big( \log{ \hat{p}\big(k | \mathbf{RS}_{a,b,c};(\mathbf{X}^i,\Theta^{i})\big) } \Big) \cdot  p\big(k | \text{IP}_{a,b,c}^{Real};(\mathbf{X}^i)\big) 
\end{equation}

\autoref{ParticularLossFunction} represents the sum of all CE in IP between the predicted conditional probability and the real probability $p\big(k)$ from IP$^{Real}$. Any location $(a,b,c)$, whose category in IP$^{Real}$ is $k$, has a real probability vector of $[0.. \,1\, ..0]$ with $1$ at the $k$-position, so \autoref{ParticularLossFunction} is highly simplified when calculated. 

Each CNN$_i$ receives a mini-batch of $m$ samples from DB$^i$ and estimates the gradient with respect to the loss function, $\nabla_{\Theta} \mathcal{L}_i(\Theta^{i}) \approx \nabla_{\Theta} [\nicefrac{1}{m} \sum_{k=1}^m \mathcal{L}_i\big (\Theta^{i}; \mathbf{X}_{k}^{i}\big )]$, and performs a parameter update $\Theta_t^{i} \gets \Theta_t^{i} + \triangle \Theta^{i}$ by inferring the updated direction $\triangle \Theta^{i}$ with respect to the gradient $\nabla_{\Theta} \mathcal{L}_i \big (\Theta^{i} \big)$ by using the Adam Optimizer \citep{kingma2014adam}. After all CNN$_i$ have been trained by all mini-batches, the first epoch is completed. The entire process is carried out again as many times as the number of epochs previously defined, or until the entire network shows signs of convergence. 

The \textsc{RCNN} simulation process begins by migrating the conditioning data to the closest node at each $D^i$. First $D^1$ is fully simulated, then $D^2$ until $D^N$ is completely informed. The sequence of nodes to be simulated at each $D$ is given by the same random path, previously defined. Following that path and at unknown locations over $D^i$, the collocated SG associated to $D^{i-1}, D^{i-2},..D^0$ are extracted, concatenated and fed into the CNN$_i$ to obtain the IP$^i$. Then, instead of freezing all IP$^i$ values in D$^i$, only a random percentage of them are selected and frozen at unknown locations. Particularly, the unknown center is always simulated. The percentage of random values used across this paper is 50\%.

\section{Experiments} \label{exp}

\begin{figure}[H]
\begin{center} 
\includegraphics[width=.95\textwidth]{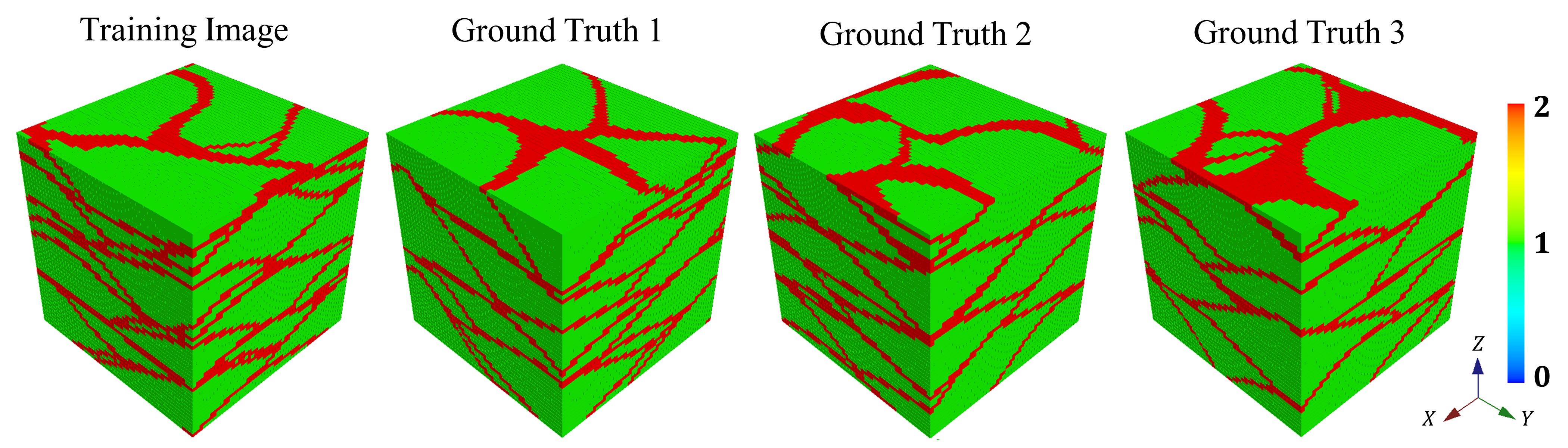}
\caption{Training image and three ground truth. All of them with sizes $50 \times 50 \times 50$.}
\label{figTI_and_GT}
\end{center}
\end{figure}

A three dimensional binary lithological surface-based model is synthetically built with an extension of $100 \times 100 \times 50$ pixels. Categories are indexed as ($s_0: 0$) unknown category, ($s_1: 1$) category 1 and ($s_2: 2$) category 2. The field is split in 4 sectors of $50 \times 50 \times 50$. The first one is used as training image to train the \textsc{RCNN} and the other three as ground truths, named $S1$, $S2$ and $S3$, as shown in \autoref{figTI_and_GT}. 

\begin{figure}[H]
\begin{center} 
\includegraphics[width=.95\textwidth]{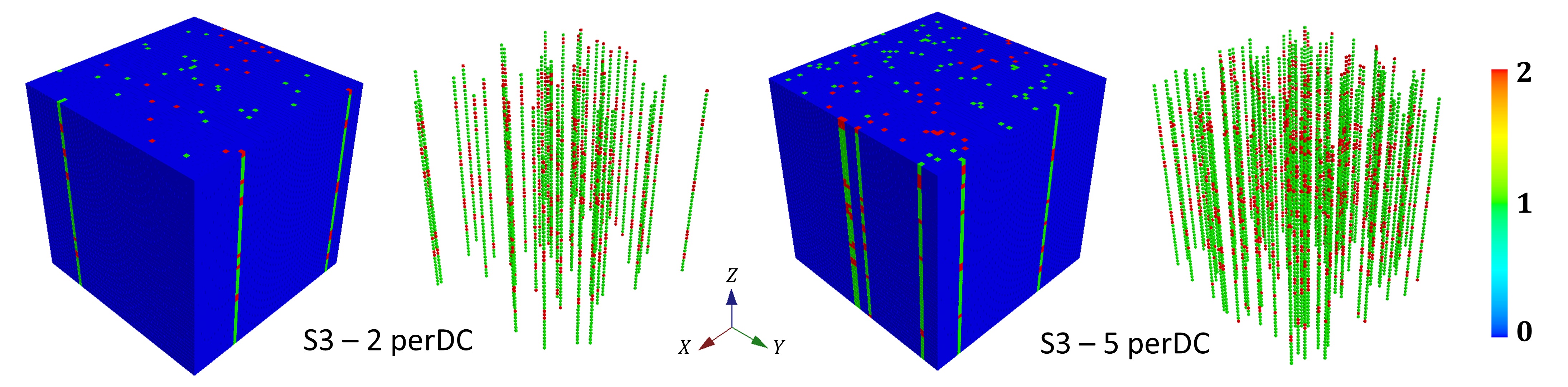}
\caption{Randomly conditioning data (drill-holes) at S3 with 2 \% (left) and 5\% (right).}
\label{figDC_Images}
\end{center}
\end{figure}

At each sector $S1$, $S2$ and $S3$ two sets of vertical drill-holes samples are randomly selected and used as conditioning data, one with $5 \%$ and the other with $2 \%$. For instance, \autoref{figDC_Images} shows the selected conditioning data over $S3$. 

The \textsc{RCNN} architecture consists of four identical nested CNNs with four hidden layers ($M^i=4$), three fully connected layers ($F^i=3$) and two categories ($K=2$). Two metrics are used to measure the quality of reproducing the spatial complexity of the underlying phenomena: 

\begin{description}
\itemsep0em
\item [\textbf{Visualizations}] One random realization, the $E$-type over a hundred realizations and the respective local variance map. 

\item [\textbf{Variograms}] Omni-horizontal and vertical experimental variograms over the TI, GT and each realization are compared, providing a measure of the two-points statistics. 
\end{description}

\section{Results} \label{secResults}
After training the RCNNs and performing 100 realizations at each of the three sectors with both sets of conditioning data, the following results are achieved. \\

\textbf{Visualizations}

From a visual point of view, the quality of lithological structure reproduction is similar between sectors. 

\begin{figure}[H]
\begin{center} 
\includegraphics[width=0.95\textwidth]{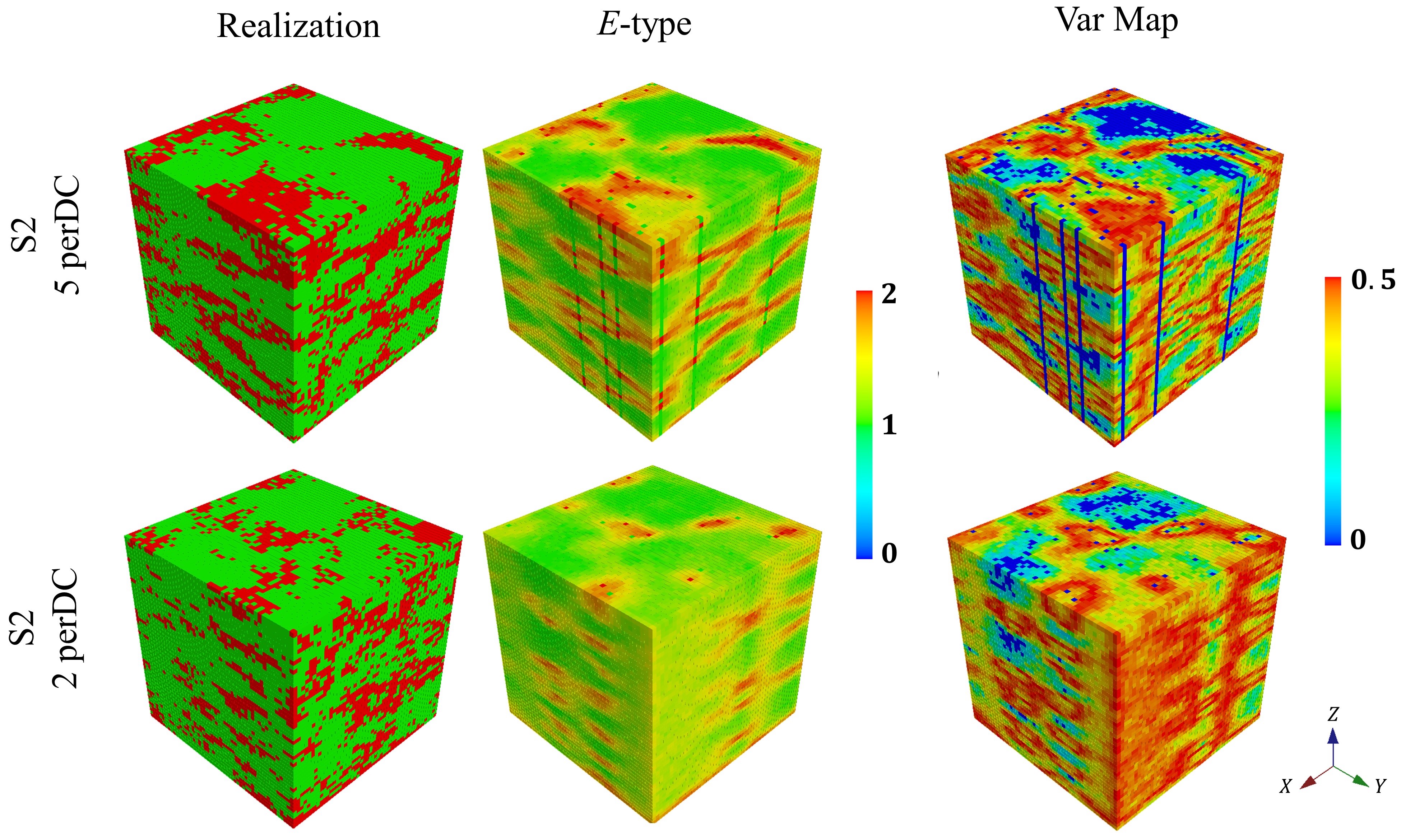}
\caption{Visualization of S2. (Left) one random realization, (center) $E$-type and (right) variance map, over 100 realizations. (Top) using 5 \% and (bottom) using 2 \% of conditioning data.}
\label{figVisualization_S2}
\end{center}
\end{figure}

\autoref{figVisualization_S2} shows visual results obtained over $S2$. Each realization honours the presence of structures with noise, connecting the conditioning data. Areas with clear absence of structure are also honoured. The $E$-type and VarMap help visualizing the predicted categories. Shapes and continuities are reasonably captured with $5\%$ of hard data but they become more diffuse with $2 \%$ of conditioning data, specially in areas larger than the search grid where there are not conditioning data.

\begin{figure}[H]
\begin{center} 
\includegraphics[width=0.9\textwidth]{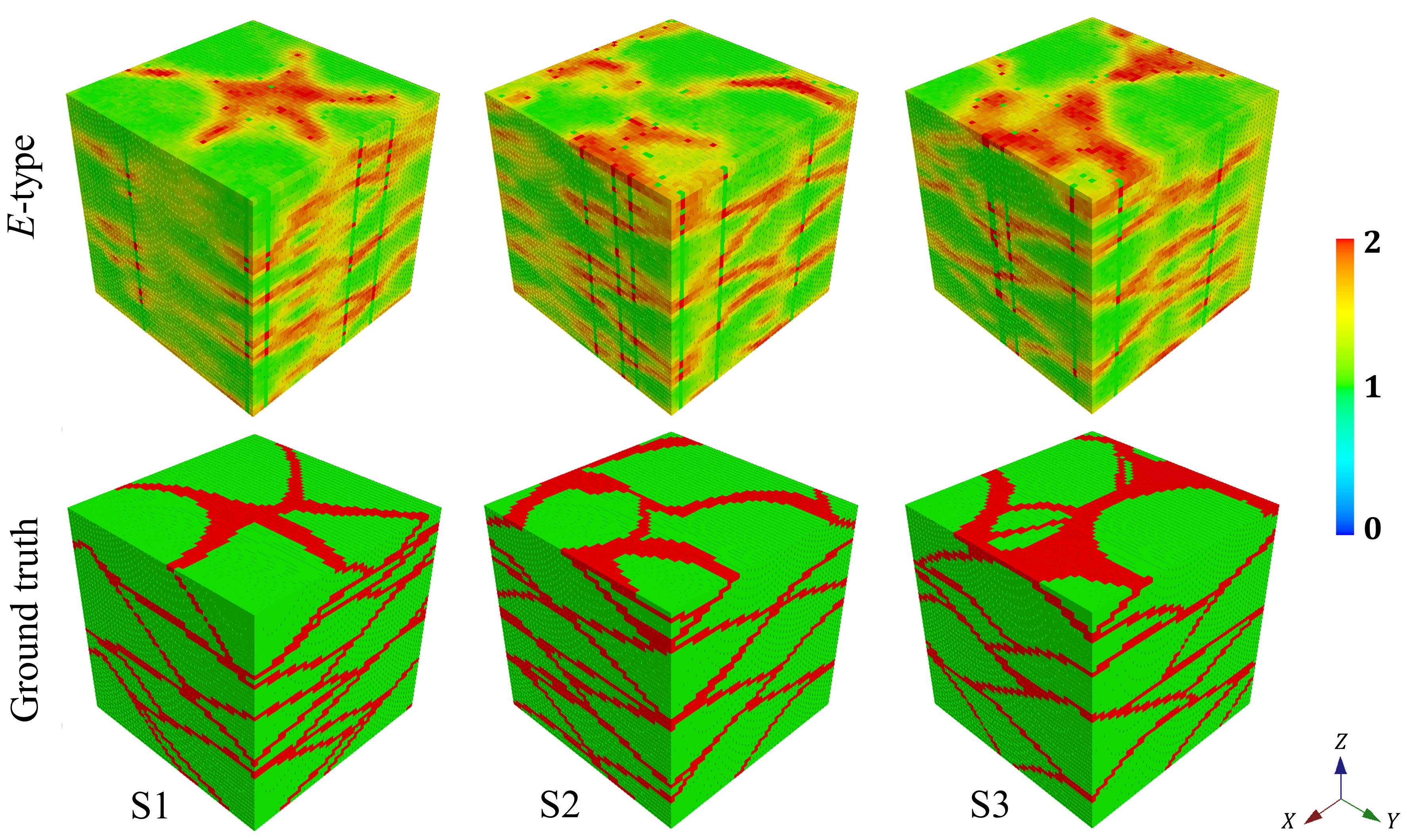}
\caption{Visualization of $E$-types (top) and Ground Truths (bottom) at $S1$ (left), $S2$ (center) and $S3$ (right) using $5 \%$ of conditioning data.}
\label{figEtypes}
\end{center}
\end{figure}

\autoref{figEtypes} shows the $E$-types of each sector using $5 \%$ of conditioning data and the associated Ground Truths. It shows the ability of the \textsc{RCNN} to predict the presence and absence of lithology at each sector honouring and using the information of the available hard data. The binary proportions are not imposed in the loss function and remain a challenge in the training of the \textsc{RCNN}. \\

\textbf{Variograms}

\begin{figure}[H]
\begin{center} 
\includegraphics[width=1\textwidth]{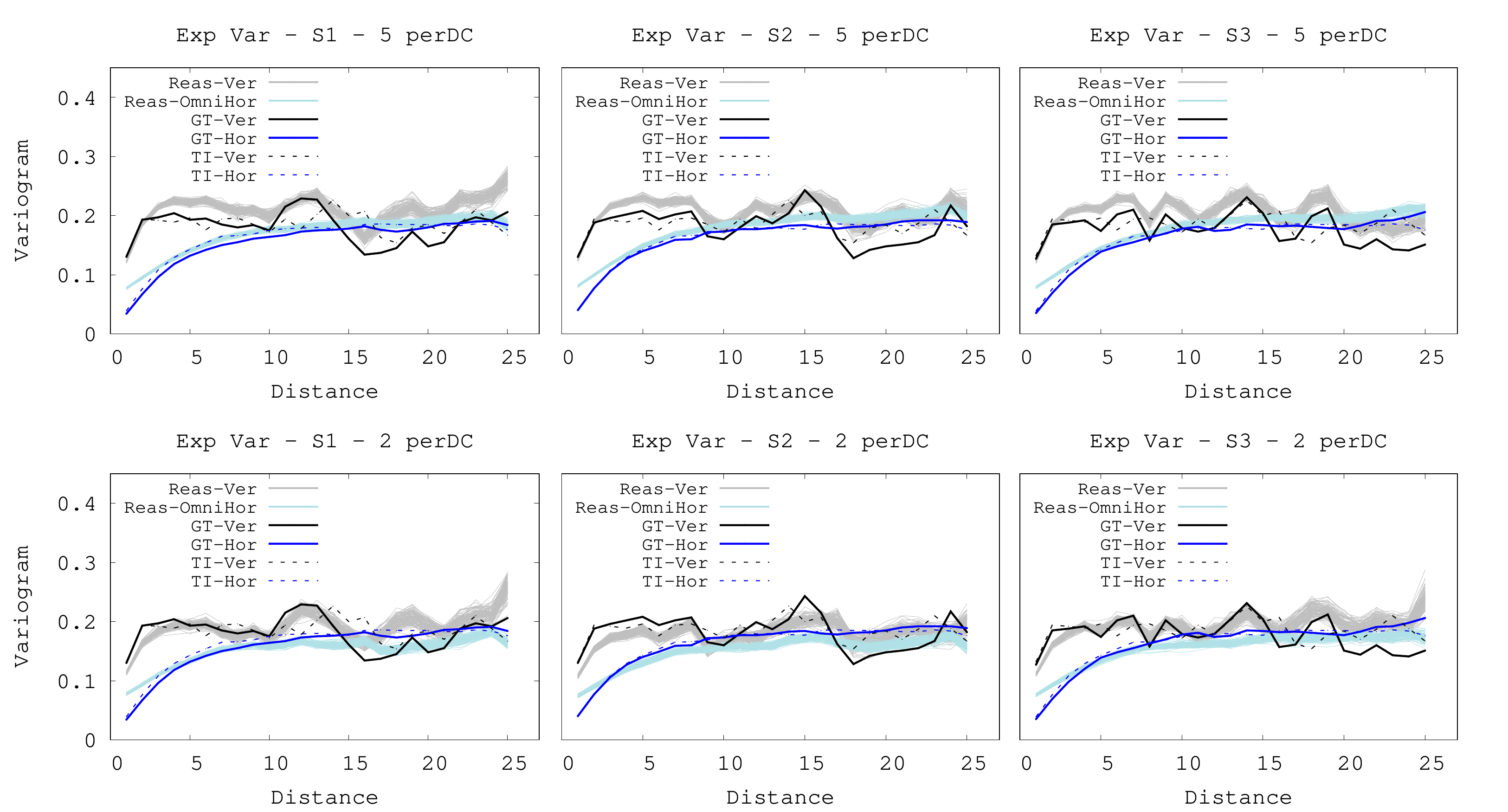}
\caption{Experimental variograms using $5 \%$ (top) and $2 \%$ (bottom) of hard data at S1 (left), S2 (center) and S3 (right).}
\label{figVariograms_Res}
\end{center}
\end{figure}

From a variographic perspective (\autoref{figVariograms_Res}), the $\textsc{RCNN}$ is able to reproduce the two-point statistics illustrated in the shapes of realization-variograms compared with the training image and ground truths. As the distance increases, variograms from realizations fluctuate as the GT does rather than the fluctuations of TI-variogram. Also, reasonable variance are reached by the realization-variograms: $(0.17, 0.19)$ with $2\%$ and $5\%$ of conditioning data respectively, compared with the variance of $0.18$ found in the TI/GT.

\section{Conclusions} \label{ConclSec}

One of the hardest issues to accomplish by MPS techniques is their implementation and effectiveness in three dimensions. During the present work, the methodology of a 3D implementation of the proposed \textsc{RCNN} technique was presented, applying it into a lithological case where good results were achieved and measured, showing the benefits of bringing Deep Learning techniques in the MPS framework. 

By means of visualizations and variograms the ability of \textsc{RCNN} to learn the spatial arrangement of lithological structures, by training, and the reproduction of it, by simulation, was displayed, measured and interpreted. Shapes and connectivities were captured using $5 \%$ and $2 \%$ of hard data.\\

\begin{small}
\noindent \textbf{Acknowledgments}. The authors acknowledge the funding provided by the Natural Sciences and Engineering Council of Canada (NSERC), funding reference number RGPIN-2017-04200 and RGPAS-2017-507956, and the Computer and Geoscience Research Scholarship 2018. Special thanks to Honggeun Jo and Prof. Michael J. Pyrcz from the Texas Center for Geostatistics for providing the training image.
\end{small}

%\section*{References}

\bibliography{Bibliography}

\end{document}